\documentclass[english,twocolumn]{revtex4}
\usepackage[T1]{fontenc}
\usepackage[latin1]{inputenc}
\usepackage{float}
\usepackage{graphicx}

\makeatletter

\usepackage{graphicx}
\usepackage{color}

\usepackage{babel}
\makeatother
\begin{document}

\title{Exact Solution of Ising Model on a Small-World Network}

\author{J. Viana Lopes, Yu. G. Pogorelov, J. M. B. Lopes dos Santos }

\address{Centro de Física do Porto, Departamento de Física, Faculdade de Ciências,
Universidade do Porto, 4169-007 Porto Portugal}

\author{and R. Toral}

\address{Instituto Mediterraneo de Estudios Avanzados IMEDEA (CSIC-UIB). Ed.
Mateu Orfila, Campus E-07122 Palma de Mallorca (Spain).}

\begin{abstract}
We present an exact solution of a one-dimensional Ising chain with
both nearest neighbor and random long-range interactions. Not surprisingly,
the solution confirms the mean field character of the transition.
This solution also predicts the finite-size scaling that we observe
in numerical simulations.
\end{abstract}
\maketitle

\section{Introduction}

Physical interactions are usually of limited range. Nevertheless,
there is a long history of study of the effect of infinite range interactions
in basic models of statistical physics \cite{baker63,nagle70}. More
recently, the work of Watts and Strogatz \cite{watts_strogatz} on
small world networks brought renewed attention to this question. Watts
and Strogatz considered a quantitative model that interpolates between
a regular lattice and a random one. They showed that the addition
of a small fraction of random long range links can dramatically change
the connectivity properties of the lattice. In particular, in a lattice
of $N$ nodes the averaged chemical distance, that is the averaged
minimum number of links between any two nodes, is of order $\mathcal{O}(\ln N)$,
rather than $\mathcal{O}(N)$, as in a regular lattice, regardless
of the number $pN$ of additional links ($p>0$), provided $N$ is
large enough. Clearly, this change should be reflected in the phenomenology
of any physical models defined on such lattices. Researchers were
quick to grasp this opportunity and many physical models and processes
have been considered in small-world networks, like cellular automata
\cite{watts99}, diffusion \cite{monasson99}, neural networks \cite{fernandez00},
the spread of disease \cite{newman99}, and many others \cite{newman00}. 

The Ising model, the simplest paradigm of order-disorder transitions,
has been studied on small-world networks in one dimension (1D) both
numerically and analytically \cite{dorogovtsev03,barrat00,gitterman00,hong02},
and in 2D and 3D numerically \cite{herrero02}. Some partial analytic
results, based on the use of the replica trick, were presented by
Gitterman \cite{gitterman00} and Barratt and Weigt \cite{barrat00}.
They are, however, contradictory (Gitterman predicts a ferromagnetic
phase only for $p\ge1/2$) and, at any rate, these results do not
constitute a full solution of the thermodynamics of the model. There
is a consensus that the ferromagnetic Ising model has a mean-field
transition for any finite value of $p$ even in 1D, even though Hastings
predicts the occurrence of anomalous scaling of the mean-field amplitudes
with $p\to0$ \cite{hastings03}. Dorogovtsev \emph{et al} \cite{dorogovtsev02}
presented a very general discussion, for networks that are locally
tree-like, based on the solution of the Ising model on a Bethe lattice
\cite{baxter82}. 

In this work we present an explicit solution of the 1D Ising model
with additional random long-range bonds. We are able to calculate
not only the transition temperature, but also the complete thermodynamics
at all temperatures as a function of $p$ and of the strength of long
range interactions $I$ (though in this paper we only present the
results for the free energy and specific heat, in zero field). The
form of finite size scaling corrections to the thermodynamics can
also be derived from our solution. 

The following Sec. \ref{sec:The-Model} presents a description of
our specific model and the basic algorithm for its exact solution.
The combinatorial treatment of its constituents is given is Sec. \ref{sec:Combinatorics-of-bonds}.
In Sec. \ref{sec:Thermodynamic-Behaviour} we analyze the resulting
thermodynamics and conclude on the mean-field character of the ordering
transition. Sec. \ref{sec:Finite-Size-scaling} shows how the finite
size scaling properties can be obtained within the same approach.
Finally, the relation between our results and those of Dorogovtsev
\emph{et al} is discussed in Sec. \ref{sec:Relation-to-Bethe}.

\section{The Model}

\label{sec:The-Model}Watts and Strogatz originally considered a model
in which the bonds of a regular lattice are rewired at random with
a probability $p$. It is widely believed that the modified model,
in which random long range bonds, or \emph{shortcuts,} are added to
the regular lattice \cite{newman99}, is essentially equivalent. 

The actual model consists of a chain of $N$ Ising spins $\sigma_{i}$,
with nearest neighbor interactions $J$ (chain bonds) and shortcut
interactions $I$ (long range bonds), both $J$ and $I$ being positive,
so that the Hamiltonian reads \begin{equation}
\mathcal{H}=-J\sum_{i=0}^{N-1}\sigma_{i}\sigma_{i+1}-I\sum_{(ij)\in S}\sigma_{i}\sigma_{j}-h\sum_{i=0}^{N-1}\sigma_{i}\label{eq:1}\end{equation}
with $\sigma_{i}=\pm1$ and periodic boundary conditions: $\sigma_{0}=\sigma_{N}$.
The set $S$ contains $N_{b}=pN$ shortcut pairs of spins, and the
last term accounts for the effect of external magnetic field. We present
in detail the solution of a version of this model in which \emph{}the
sites connected by shortcuts are equally spaced in the regular lattice,
a distance $1/2p$ apart, but the shortcuts are randomly arranged
among these spins. The solution of theoriginal model, in which the
$2pN$ sites are randomly distributed along the lattice, turns out
to be essentially the same, and some results for this latter case
are also presented in the Appendix. In either case, the bond selection
is such that no spin is linked to more than one shortcut. 

We consider a transformation from site variables to bond variables,
which we formulate for a general Ising model in zero magnetic field,\begin{equation}
\mathcal{H}=-\sum_{(i,j)}J_{ij}\sigma_{i}\sigma_{j},\label{eq:morto1}\end{equation}
where $J_{ij}$ can be chosen arbitrary (though in what follows they
are only $0,$ $J$ or $I$). The partition function for temperature
$T=\beta^{-1}$ is given by\begin{eqnarray}
\mathcal{Z} & = & \textrm{Tr}_{\{\sigma\}}\exp(\beta\sum_{(i,j)}J_{ij}\sigma_{i}\sigma_{j})\label{eq:morto2}\\
 &  & =\textrm{Tr}_{\{\sigma\}}\prod_{(i,j)}\exp(\beta J_{ij}\sigma_{i}\sigma_{j}).\nonumber \end{eqnarray}
We can use the known identity \begin{equation}
\exp(\beta J_{ij}\sigma_{i}\sigma_{j})=\cosh\beta J_{ij}\left(1+\sigma_{i}\sigma_{j}\tanh\beta J_{ij}\right)\label{eq:morto3}\end{equation}
and then represent the latter factor as a sum in a discrete {}``bond
variable'' $b_{ij}=0,1$:\begin{equation}
1+\sigma_{i}\sigma_{j}\tanh\beta J_{ij}=\sum_{b_{ij}=0,1}\left(\sigma_{i}\sigma_{j}\tanh\beta J_{ij}\right)^{b_{ij}},\label{eq:morto4}\end{equation}
to write the partition function as\begin{eqnarray}
\mathcal{Z} & = & \left(\prod_{(i,j)}\cosh\beta J_{ij}\right)\label{eq:morto4}\\
 &  & \times\textrm{Tr}_{\{\sigma\}}\sum_{\{ b\}}\prod_{(i,j)}\left(\sigma_{i}\sigma_{j}\tanh\beta J_{ij}\right)^{b_{ij}}.\label{eq:morto5}\end{eqnarray}
When we trace over any spin variable $\sigma_{i}$, in a product with
fixed configuration $\{ b\}$ of bond variables $b_{ij}$, we get
zero if the $i$th spin multiplicity in this product, $\sum_{j}b_{ij}$,
is odd and a factor of 2 if it is even. Therefore we can trace over
the spin variables to obtain \begin{eqnarray}
\mathcal{Z} & = & 2^{N}\left(\prod_{(i,j)}\cosh\beta J_{ij}\right)\\
 &  & \times\sum_{\{ b\}}\prod_{(i,j)}\left(\tanh\beta J_{ij}\right)^{b_{ij}},\label{eq:2}\end{eqnarray}
where the sum over $\{ b\}$ is restricted to configurations with
only even $\sum_{j}b_{ij}$ for all $i$. 

As an example, we can derive from Eq.~(\ref{eq:2}) the partition
function of the Ising chain ($J_{ij}=J$, for nearest neighbors).
Since each spin there has only two bonds, this implies that either
$b_{ij}=0$ for all the bonds or $b_{ij}=1$ for all the bonds, and
these two alternatives contribute in the partition function\begin{equation}
\mathcal{Z}_{chain}=(2\cosh\beta J)^{N}\left(1+\tanh^{N}\beta J\right)\label{eq:morto7}\end{equation}
as expected.

The restriction of the sum in Eq.~(\ref{eq:2}) to configurations
such that $\sum_{j}b_{ij}$ is even, allows a reduction of the number
of required variables. If the site $i$ has no shorcut bonds, then
this condition reduces to $b_{i-1,i}+b_{i,i+1}$ being an even number,
that is, $b_{i-1,i}=b_{i,i+1}$ (since $b_{ij}=0,1$). On the other
hand, if $i$ is a shorcut site, with an associated additional bond
variable, the sum in Eq.~(\ref{eq:2}) gets only nonvanishing contributions
when $b_{i-1,i}=b_{i,i+1}$, if the additional bond variable is $0$,
or when $b_{i-1,i}=1-b_{i,i+1}$, if the additional bond variable
is $1$. In any case, the knowledge of one chain bond variable (say
$b_{0,1}$) and the shorcut variables is enough to determine all the
terms that give a nonzero contribution to Eq.~(\ref{eq:2}). Hence,
the partition function can be written as an unrestricted sum over
$b_{0,1}$ and $N_{b}$ shorcut bond variables which, in order to
simplify notation, we denote simply by $b_{0},b_{1},\dots,b_{N_{b}}$:

\begin{equation}
\mathcal{Z}=\mathcal{Z}_{chain}c_{I}^{pN}\sum_{\left\{ b_{0},\dots,b_{N_{b}}\right\} }t_{J}^{L[b]}t_{I}^{M[b]}\label{eq:3}\end{equation}
where $c_{I}\equiv\cosh\beta I$, $t_{J}\equiv\tanh\beta J$, $t_{I}\equiv\tanh\beta I$,
$M=\sum_{i=1}^{n}b_{i}$ is the number of shortcut bonds with $b=1$,
and $L$ the number of chain bonds with $b=1$. We calculate it as
follows: given a configuration of $N_{b}$ shortcuts and a choice
of $M$ from them with $b_{i}=1$, the chain gets divided into $2M$
segments between consecutive spins connected to one of $M$ bonds.
These segments have lengths $l_{1},\dots,l_{2M}$. When $b_{0}=0$,
the value $L[b]$ is the sum of even lengths, $l_{2}+l_{4}+\dots+l_{2M}$.
Otherwise, it is the sum of the odd ones, $l_{1}+l_{3}+\dots+l_{2M-1}$. 

Our solution is based on the observation that Eq.~(\ref{eq:3}) permits
explicit separation of two extensive thermodynamical variables $L[b]$
and $M[b]$ (both being $\mathcal{O}(N)$) from $N_{b}$ microscopic
variables $b_{i}$. The latter only define a temperature independent
prefactor $\Omega(M,L)$ at the product $t_{J}^{L}t_{I}^{M}$ with
given $L$ and $M$. A similar situation (but with a single variable
$t_{J}^{L}$) was already addressed by Saul and Kardar for the problem
of Ising spin glasses \cite{saul}. Instead of numerical procedure
for the prefactor employed in Ref. \cite{saul}, we are able here
to calculate analytically the sum over all the bond configurations
in Eq.~(\ref{eq:3}). Denoting $\Omega(M,L)=\exp[S(M,L)]$ the number
of choices of $M$ segments with a fixed value of $L$, we get\begin{equation}
\mathcal{Z}=\mathcal{Z}_{chain}c_{I}^{pN}\sum_{b_{0}}\sum_{M,L}\exp(S-L/\xi_{J}-M/\xi_{I}),\label{eq:4}\end{equation}
where $\xi_{J}=1/\ln\left(1/t_{J}\right)$ is the Ising chain correlation
radius and $\xi_{I}=1/\ln\left(1/t_{I}\right)$. 

The crucial property of the exponential function under sum in Eq.~(\ref{eq:4})
is that it has, as will be seen below, an extremely sharp maximum
at some \emph{macroscopically} great values of $L$ and $M$. This
reduces the calculation of its contribution to the free energy, $\mathcal{F}=-T\ln\mathcal{Z}$,
to the logarithm of its maximum. The latter turns out to be $\mathcal{O}(N)$
and thus a self-averaging quantity, that is coinciding, for (almost)
any random realization of the disordered system, with its average
value \cite{lifshitz}. 

Passing to the intensive thermodynamical variables $l\equiv L/N,$
$n\equiv M/N_{b}$, and to the intensive function $s(l,n)=S(L,M)/N$,
we rewrite Eq.~(\ref{eq:4}) as\begin{equation}
\mathcal{Z}=2\mathcal{Z}_{chain}c_{I}^{pN}\sum_{l,n}\exp[N\left(s-l/\xi_{J}-pn/\xi_{I}\right)],\label{eq:5}\end{equation}
and in the thermodynamic limit $N\to\infty$ the sum converts into
integral which can be done by steepest descent. The values $b_{0}=0$
and $1$ give identical contributions, hence the factor 2. The free
energy per spin is $f=\mathcal{F}/N=-(T/N)\ln\mathcal{Z}=-(T/N)\ln\mathcal{Z}_{0}\mathcal{Z}_{a}=f_{0}+f_{a}$,
where \begin{equation}
f_{0}=-T[\ln(2\cosh\beta J)+p\ln\cosh\beta I]\label{eq:6}\end{equation}
and the {}``anomalous'' term $f_{a}$ (if exists, see below) is
given by simple minimization with respect to $l$ and $n$ of the
following function \begin{eqnarray}
f(l,n) & = & T[-s(l,n)+l/\xi_{J}+pn/\xi_{I}].\label{eq:7}\end{eqnarray}
The task that remains is to calculate the configurational entropy
$s(l,n)\equiv\ln\Omega(L,M)/N$ (for $L,M=\mathcal{O}(N)$).

\section{Combinatorics of bonds }

\label{sec:Combinatorics-of-bonds}The calculation of $\Omega(L,M)$
can be formulated in the following way. 

\begin{figure}[H]
\begin{center}\includegraphics[%
  width=0.90\linewidth,
  keepaspectratio]{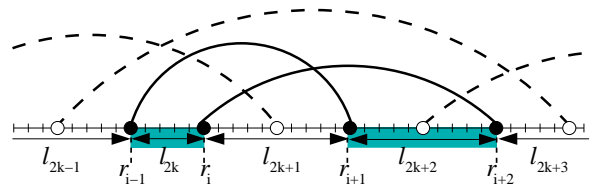}\end{center}

\caption{\label{Esquema_links} $\Omega(L,M)$ is the number of possible choices
of $M$ bonds, terminating in filled dots, from $pN$ fixed bonds,
which have a given value of the sum $l_{2}+l_{4}+\dots+l_{2M}=L$. }
\end{figure}

We are given a chain of $N$ sites $\{0,1,2,\dots,N-1\}$ with periodic
boundary conditions. The shortcut sites are evenly spaced forming
a regular lattice with coordinates $\{ d,2d,\dots,N\}$, where $d=1/2p$.
A number $pN$ of shortcuts connect $pN$ pairs, randomly chosen from
these sites. If we choose $M$ of these bonds (those for which $b=1$)
from the total of $pN$, the corresponding shortcut sites (the filled
dots in Fig.~(\ref{Esquema_links})) will have coordinates, in increasing
order $\{ r_{1},r_{2},\dots,r_{2M}\}\subseteq\{ d,2d,\dots,N\}$.
These site coordinates will divide the lattice into $2M$ segments
of lengths $dl_{1},\dots,dl_{2M}$ where \begin{eqnarray}
l_{i} & = & \frac{r_{i}-r_{i-1}}{d},\qquad i\neq1,\label{eq:morto8}\\
l_{1} & = & \frac{N-r_{2M}+r_{1}}{d},\nonumber \end{eqnarray}
and $\sum_{i=1}^{2M}l_{i}=N/d$. Then $\Omega(L,M)$ is the number
of possible choices of the $M$ bonds such that \begin{equation}
l_{2}+l_{4}+l_{6}+\dots+l_{2M}=L/d.\label{eq:8}\end{equation}

We are asking in how many ways one can divide the interval $[0,N/d]$
into $2M+1$ integer lengths, $\{ l_{1},\dots,l_{2M+1}\}$, where
only $l_{2M+1}$ may be zero, so that even lengths are summed to\begin{equation}
l_{2}+l_{4}+\dots+l_{2M}=L/d.\label{eq:morto9}\end{equation}
Then, clearly, the odd lengths should sum to\begin{equation}
l_{1}+l_{3}+\dots+l_{2M+1}=\frac{N-L}{d}.\label{eq:morto10}\end{equation}
Therefore $\Omega(L,M)$ is just the number of ways of dividing the
interval $[0,L/d]$ into $M$ segments of integer length, times the
number of ways of dividing $[0,(N-L)/d]$ into $M+1$ integer segments
(the last of them possible to be zero): \begin{equation}
C_{M}^{(N-L)/d}C_{M-1}^{L/d-1}\label{eq:morto11}\end{equation}
(where $C_{m}^{n}=n!/[m!(n-m)!]$, a binomial coefficient). 

Note, however, that our choice of $2M$ coordinates $\{ r_{1},\dots,r_{2M}\}$
is constrained by the the fact that the original problem has \emph{bonds}
connecting pairs of sites $q_{k},$$q_{m}$. Therefore, if $q_{k}\in\{ r_{1},\dots,r_{2M}\}$
so must $q_{m}$. To take this fact into account, we multiply the
previous factor by the normalizing factor, \begin{equation}
C_{M}^{pN}/C_{2M}^{2pN},\label{eq:morto12}\end{equation}
which refers to the self-averaging property in the thermodynamic limit:
that (almost) all possible pairings between the sites $\{ d,2d,\dots,N\}$
give the same $\Omega(L,M)$. Considering Eq.~(\ref{eq:3}), we express

\begin{eqnarray}
\Omega(L,M) & = & C_{M}^{2p(N-L)}C_{M-1}^{2pL-1}C_{M}^{pN}/C_{2M}^{2pN}\label{eq:morto13}\end{eqnarray}

\begin{figure}
\begin{center}\includegraphics[%
  width=0.80\linewidth,
  keepaspectratio]{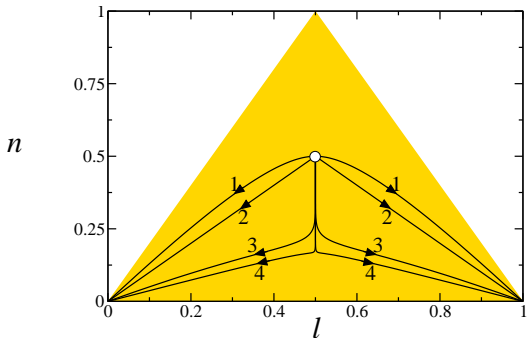}\end{center}

\caption{\label{trajectoria}Trajectories of the maxima of the exponent in
Eq.(\ref{eq:5}) in the domain $\{ l,n\}$, as temperature varies
from $T=0$ (central point), to $T_{c}$ (lateral vertices, $b_{0}=0$
or $b_{0}=1$) at particular choices of parameters: 1) $p=0.01$,
$I=1$; 2) $p=1/2$, $I=1$; 3) $p=0.1$, $I=0.1$; 4) $p=1/2$, $I=0.1$.
For each temperature there are two equal contributions to free energy
resulting from $b_{0}=1,0$. }
\end{figure}

At this point it is important to specify the variation range for the
variables $L$ and M. Clearly, $0<M<pN$. On the other hand, since
each $r_{i}-r_{i-1}$ measures at least $d=1/2p$, we must have $2pL\ge M$
and $pN-2pL\ge M$, i.e\emph{.} the before defined intensive variables
$n=M/pN$ and $l=L/N$ should belong to the triangle\emph{\begin{equation}
n\le2l,\quad\quad n\le1-2l\qquad0\le n\le1,\label{eq:9}\end{equation}
}shown in Fig.~(\ref{trajectoria}). 

For our purposes, all that will be required is the leading $\mathcal{O}(N)$
term in $\ln\Omega(L,M)$. Using Stirling's formula\begin{equation}
n!\approx\sqrt{2\pi n}\left(\frac{n}{e}\right)^{n},\label{eq:morto01}\end{equation}
we arrive at\begin{equation}
\ln\Omega(L,M)=Ns(l,n)\label{eq:morto02}\end{equation}
with \begin{eqnarray}
s(l,n) & = & p\left[\left(2-2l\right)\ln\left(2-2l\right)+2l\ln(2l)\right.\nonumber \\
 &  & +(1-n)\ln(1-n)-n\ln n\nonumber \\
 &  & -\left(2l-n\right)\ln\left(2l-n\right)\nonumber \\
 &  & \left.-\left(2-2l-n\right)\ln\left(2-2l-n\right)\right],\label{eq:10}\end{eqnarray}
(dropping the terms $O(\ln N$)). 

In order to confirm the basic assumption of self-averaging, involved
in the calculations leading to Eq.~(\ref{eq:10}), we performed multicanonical
Monte Carlo simulations on the variables $l$ and $n$. For the sampling
probability distribution of a state $i$ (characterized by a set of
values for the variables $\{ b\}$) we used $p_{i}=1/\Omega(L_{i},M_{i})$.
The Monte Carlo move is given by:

\begin{itemize}
\item From an initial state $i$, propose a new state $j$, differing from
$i$ by the value of a randomly chosen bond (with probability $1/N_{b}$).
\item Accept the new state with the usual probability $W_{ij}=\min(1,\Omega(L_{i},M_{i})/\Omega(L_{j},M_{j}))$.
\end{itemize}
This procedure ensures (by detailed balance and ergodicity of the
algorithm) that the asymptotical sampling probability is $p_{i}$.
The asymptotical histogram of frequencies in an $(l,n)$ point should,
then, be given by,\[
H(l,n)\propto\Omega_{\alpha}(L,M)\frac{1}{\Omega(L,M)}\]
where $\Omega_{\alpha}(L,M)$ is the actual number of states, with
given $(L,M)$, of the sample that we are using in the simulation.
If $\Omega_{\alpha}(L,M)=\Omega(L,M)$, the histogram should be flat
in all the range of $(l,n)$.

\begin{figure}
\begin{center}\includegraphics[%
  width=0.70\linewidth]{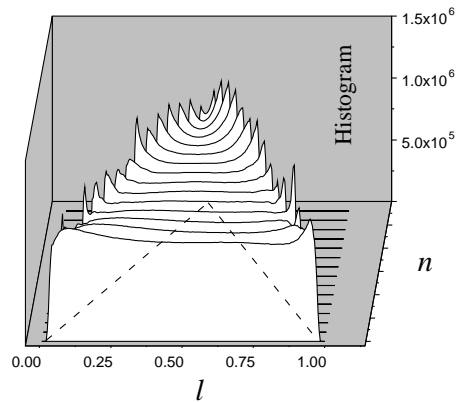}\end{center}

\caption{\label{cap:Histogram} Histogram of occurrences of $l$ values for
several fixed values of $n$, obtained in multicanonical simulations
with the entropy of Eq.~(\ref{eq:10}) as sampling distribution for
a sample with $p=1/4$ and $N=512$. }
\end{figure}

The corresponding histograms (shown in (Fig.~(\ref{cap:Histogram}),
for a single disorder realization) are indeed quite flat , confirming
that the entropies we calculated are apparently exact in the thermodynamic
limit and very accurate for the moderate tested sizes. Recall that
$\Omega(L,M)\propto\exp(Ns(l,n))$ and, for large $N$, even small
deviations of $\ln\Omega_{\alpha}(L,M)$ from $S(l,n)$, would result
in simulations witch do not cover the entire spectrum, as ours do. 

\section{Thermodynamic Behaviour}

\label{sec:Thermodynamic-Behaviour}The necessary conditions for that
the exponent in Eq.~(\ref{eq:5}) is maximum: $\partial f(l,n)/\partial n=\partial f(l,n)/\partial l=0$,
define from Eq.~(\ref{eq:10}) the following equations for the coordinates
$n^{*}$ and $l^{*}$ of this maximum\begin{eqnarray}
\frac{n^{*}(1-n^{*})}{(2-2l^{*}-n^{*})(2l^{*}-n^{*})} & = & t_{I}\label{eq:mortoa}\\
\frac{(2l^{*}-n^{*})(1-l^{*})}{l^{*}(2-2l^{*}-n^{*})} & = & t_{J}^{d}\nonumber \end{eqnarray}
($d\equiv1/2p$). These equations are easily solved: \begin{eqnarray}
l^{*} & = & \frac{1-t_{J}^{d}-2t_{I}t_{J}^{d}}{(1-t_{J}^{d})^{2}-4t_{I}t_{J}^{d}},\label{eq:11}\\
n^{*} & =l^{*} & \left(1+\frac{t_{I}-t_{J}^{d}}{1+t_{I}}\right)\label{eq:12}\end{eqnarray}
and, for $t_{I},$ $t_{J}$ varying with temperature, these are parametric
equations for a certain trajectory of the maximum of the exponent
in the $n,l$ plane (within the admissible range, Eq.~(\ref{eq:9})).
At zero temperature, when $t_{I}=t_{J}=1$, it locates at $l^{*}=1/2,$
$n^{*}=1/2$ (the central point in Fig.~\ref{trajectoria}), corresponding
to the maximum configurational entropy $s(l,n)$. As $T$ increases,
the maximum moves towards the origin $n=l=0$ (or to $n=0,\: l=1$),
as shown in Fig.~(\ref{trajectoria}). The trajectory reaches the
very origin at a finite temperature $T=T_{c}$, given by the condition\begin{equation}
t_{J}^{d}(1+2t_{I})=1.\label{eq:13}\end{equation}
It is important to notice that below this critical temperature, when
the reduced temperature $t\equiv T/T_{c}-1$ is non zero ($t<0$),
no matter how small $|t|$ is, the coordinates of maximum $(l^{*},n^{*})$
Eqs.~(\ref{eq:11}, \ref{eq:12}), are also finite: $l^{*},n^{*}\sim|t|$.
Hence the corresponding numbers $L^{*}=l^{*}N$ and $M^{*}=pn^{*}N$
are macroscopic: $L^{*},M^{*}=\mathcal{O}(N)$. On the other hand,
the width of the maximum, estimated from the second derivatives $\left.\partial^{2}s/\partial l^{2}\right|_{l^{*},n^{*}}\sim\left.\partial^{2}s/\partial n^{2}\right|_{l^{*},n^{*}}\sim1/l^{*}$,
is $\mathcal{O}(\sqrt{l^{*}/N})$ and tends to zero in the thermodynamic
limit, that is the probability distribution in macroscopic variables
$l,n$ tends to a $\delta$-function with an amplitude which can only
depend on macroscopic parameters of long range links ($p$ and $I$)
but not on their specific realization. This justifies the above used
assumption of self-averaging for the probability distribution and
relates it to the known self-averaging property for observable values
\cite{lifshitz}.

Above $T_{c}$, the maximum of the exponent in Eq.~(\ref{eq:5})
goes away from the physical region, Eq.~(\ref{eq:9}), while its
highest value in this region is zero, attained at the origin. Hence
the last factor of the Eq.~(\ref{eq:5}) turns $\mathcal{O}(1)$,
and it gives no contribution to the free energy in the thermodynamic
limit. Therefore the free energy above $T_{c}$ is simply $f_{0}$,
Eq.~(\ref{eq:6}), but it gets an extra term below $T_{c}$: $f_{0}+f_{a}$,
where $f_{a}$ is given by Eq.~(\ref{eq:7}) with $l$ and $n$ given
by Eqs.~(\ref{eq:11}) and (\ref{eq:12}). These equations combined
provide a complete description of the thermodynamics of the model
at all temperatures in zero field. 

The dependence of transition temperature $T_{c}$, defined from Eq.~(\ref{eq:13}),
on the system parameters $p$, $J$, and $I$ can be easily analyzed
in characteristic limits. 

If shortcut bonds are much stronger than chain bonds, then for any
finite $p$ and $I\to\infty$, critical temperature tends to a finite
value: $T_{c}\to J/($arctanh$9^{-p})$. Since shortcut spin pairs
in this limit should be considered as single spins, the above value
defines also the transition temperature for a random graph with connectivity
4 made of Ising chains of length $d=1/2p$. In the limit of small
concentration, $p\ll1$, this critical temperature $T_{c}$ turns
small compared to $J$, the energy scale for Ising chain:\begin{equation}
T_{c}=\frac{2J}{\ln[1/(p\ln3)]},\label{eq:14}\end{equation}
and such $I$-independent behavior holds as well for moderate shortcut
strength (unless $I$ is too small: $I\ll J/|\ln(p\ln3)|$). At last,
in the limit where the shortcuts are much weaker than the chain bonds,
$pI\ll J$, we have within logarithmic accuracy\begin{equation}
T_{c}=\frac{2J}{\ln[T_{c}/(2pI)]}\approx\frac{2J}{\ln\{ J/[pI\ln(J/pI)]\}}.\label{eq:15}\end{equation}

\begin{figure}
\begin{center}\includegraphics[%
  width=0.80\columnwidth,
  keepaspectratio]{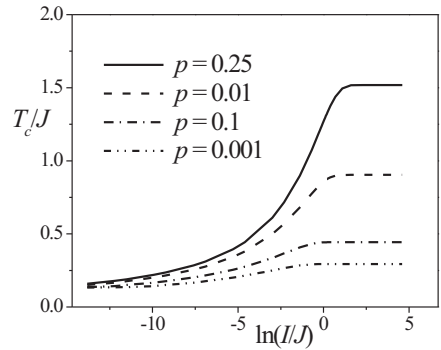}\end{center}

\caption{\label{cap:Tc}Behavior of the critical temperature (in units of
$J$) as a function of $I$ for several values of the concentration
$p$ of shortcuts}
\end{figure}
The above relations define the system phase diagram in $p,I,T$ variables,
as shown in Fig. \ref{cap:Tc}. 

It is of interest to compare these formulas to the finite critical
temperatures, resulting from breaking down the Mermin-Wagner theorem
for a 2D Heisenberg magnet in presence of (small) anisotropy $\Delta J\ll J$:
$T_{c}\approx Jz/\ln(T_{c}/\Delta J)$ ($z$ the coordination number).
They can be also referred to the percolation threshold $p_{c}$ for
the one-state limit of Potts model \cite{kasteley69}: $T_{c}=2J/\ln(1-p_{c})$.
\begin{figure}
\begin{center}\includegraphics[%
  width=0.80\linewidth]{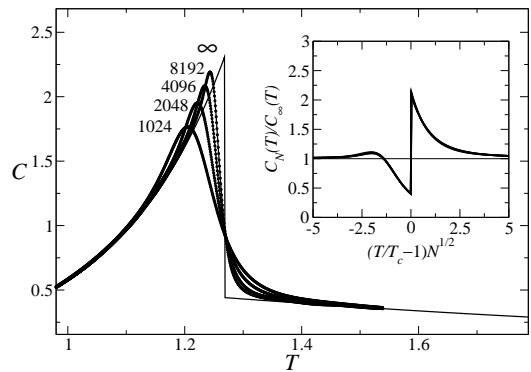}\end{center}

\caption{\label{cap:Cv}Specific heat obtained by Monte Carlo simulations
for 4 sizes of samples, with $p=1/4$ and $I=1$ in comparison with
the theoretical curve ($\infty$). Inset shows the collapse of the
4 curves when plotted in function of $tN^{1/2}$.}
\end{figure}

The non-trivial thermodynamics follows from the observation that close
to the critical point, $0<-t\ll1$, both variables $l$ and $n$ are
$\mathcal{O}(\left|t\right|)$, while the trajectory $n(l)$ reaches
the origin with asymptotic slope $dn/dl\to4/(1/t_{I}+2)<4/3$, that
is, always within the triangle, Eq.~(\ref{eq:9}). Using Eqs.~(\ref{eq:11})
and (\ref{eq:12}), the additional term $f_{a}$ in the free energy
can be simplified to: \begin{equation}
f_{a}=Tp\ln\frac{(2-2l^{*}-n^{*})^{2}}{\left(1-n^{*}\right)\left(2-2l^{*}\right)^{2}},\label{eq:16}\end{equation}
and its leading terms in the critical region are clearly of order
$\mathcal{O}(t^{2})$. Hence the specific heat per spin, $c_{v}=-T\partial^{2}f/\partial T^{2}$,
has a finite jump $\Delta C$ at the critical point. As an example
we present in Fig.~(\ref{cap:Cv}) the calculated exact specific
heat for $p=1/4$ and $I=J$, together with the results of simulations
on samples of various sizes. This behavior permits to classify the
considered transition as second order mean-field-like, with critical
exponent $\alpha=0$. In the case of $p\ll1$ with $T_{c}$ given
by Eq.~(\ref{eq:14}), the jump is proportional to $p^{3}\ln^{2}(p\ln3)$,
and is also independent of the shortcut bond strength $I$. When the
shortcuts are much weaker that the chain bonds, $pI\ll J$, with $T_{c}$
by Eq.~(\ref{eq:15}), we find the specific heat discontinuity proportional
to $p(I/J)^{2}\ln^{4}(J/pI)$. The above results indicate a non-analytic
\emph{decay} of this mean-field amplitude at $p\to0$, specific for
the considered 1D system. It is of interest to compare this to a power
law \emph{divergence} in $p$ of the mean-field amplitudes for systems
where a second order transition at non-zero temperature exists in
absence of long-range links \cite{hastings03}.

\section{Finite Size scaling}

\label{sec:Finite-Size-scaling}It was shown that the self-averaging
property invoked for our calculation of $\Omega(L,M)$ is true in
the thermodynamic limit, and flatness of the histograms in the multicanonical
simulations suggests that the calculated $\Omega(L,M)$ is accurate
(see Fig.~(\ref{cap:Histogram})). Nevertheless, there are visible
deviations from flatness near the edges of the spectrum, which diminish
with growing system size $N$. One can therefore ask whether our solution
also contains the correct finite size scaling properties of this model.

\begin{figure}
\begin{center}\includegraphics[%
  width=0.80\linewidth]{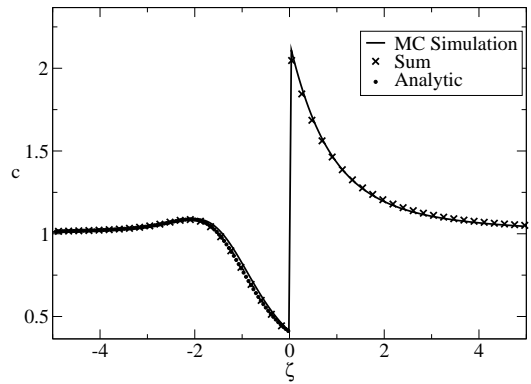}\end{center}

\caption{\label{cap:Sum-Sim} Comparison between the Monte Carlo simulation
for a sample of $N=8192$ and $p=1/4$, the numerical sum Eq. (\ref{eq:4})
and the analytical result (because of the small sizes it was necessary
to include corrections of $O(N^{-1/4})$ to the finite size scaling).}
\end{figure}

To answer this question, the numerical sum of Eq.~(\ref{eq:4}) was
performed for different temperatures. The factorials were substituted
by the Stirling's approximation and the specific heat was then obtained
by numerical differentiation. The comparison between the two independent
calculations is presented in Fig.~(\ref{cap:Sum-Sim}). Despite the
fact that the Monte Carlo simulation was made for a unique sample
(without disorder averaging), the agreement between these calculations
is very good.

To derive the analytic form of the finite size scaling function, we
approximated the sum in Eq.~(\ref{eq:4}) by an integral in the intensive
variables $l$ and $n$. Then, after the change of variables $(l,n)\to(l,u)$
where $u=n/l$, the partition function for a finite size $N$ of the
sample becomes\begin{equation}
\mathcal{Z}_{N}=\exp[-\beta N(f_{0}+f_{a})]\mathcal{Z}_{FSS}\label{eq:mortoc}\end{equation}
where\begin{equation}
\mathcal{Z}_{FSS}=\frac{N}{2\pi p}\int_{0}^{+\infty}dl\int_{0}^{2}du\, g(l,u)\exp(pN\, h(l,u))\label{eq:17}\end{equation}
defines the contributions $\sim O(N^{-1}\ln N)$ into free energy
$f$, with\begin{equation}
g(l,u)=\sqrt{\frac{2(1-l)}{(2-l(2+u))(2-u)}}\label{eq:mortod}\end{equation}
and

\begin{eqnarray}
h(l,u) & = & 2(1-l)\ln\left(\frac{1-l}{1-l^{*}}\right)+(1-ul)\ln\left(\frac{1-ul}{1-u^{*}l^{*}}\right)\nonumber \\
 & - & (2-u)l\ln\left(\frac{2-u}{2-u^{*}}\right)-ul\ln\left(\frac{u}{u^{*}}\right)\nonumber \\
 & - & \left(2-l(2+u)\right)\ln\left(\frac{2-l(2+u)}{2-l^{*}(2+u^{*})}\right).\label{eq:mortoe}\end{eqnarray}
 The function $h(l,u)$ has a maximum at $(l^{*},u^{*})$. When $T\to T_{c}^{-}$,
we have $l^{*}\to0^{+}$ and $u^{*}\to2/[1+\coth(\beta_{c}I)/2]<4/3$. 

The integral in the Eq. (\ref{eq:17}) is dominated by the vicinity
of the maximum. To obtain the leading order terms in $1/N$ and $t$
we may expand $h(l,u)$ around the maximum,\begin{equation}
h(l,u)\approx-c_{1}(l-l^{*})^{2}-c_{2}l(u-u^{*})^{2}\label{eq:mortoh}\end{equation}
with definite constants $c_{1,2}\sim O(1)$, and replace $g(l,u)$
by $g(0,u^{*})$. With suitable change of variables we obtain\begin{equation}
{\cal \mathcal{Z}_{FSS}}(\zeta)\approx\frac{N^{1/4}}{k_{1}}\int_{0}^{\infty}dx\int_{-y^{*}}^{(2/u^{*}-1)y^{*}}dy\, e^{-(x-k_{2}\zeta)^{2}-xy^{2}}\label{eq:mortog}\end{equation}
were $\zeta=tN^{1/2}$, $y^{*}\propto N^{1/4}$, and the constants
$k_{1,2}$are related to $c_{1,2}$. Since the finite size scaling
limit is $N\to\infty$ and $t\to0$ (at fixed $\zeta<0$), the limits
of integration in $y$ tend to infinity and we get

\begin{equation}
{\cal \mathcal{Z}_{FSS}}(\zeta)\approx\frac{\sqrt{\pi}N^{1/4}}{k_{1}}\int_{0}^{\infty}\frac{e^{-(x+k_{2}\zeta)^{2}}dx}{\sqrt{x}}.\label{eq:mortoi}\end{equation}
This leads to a correction in the additional free energy:\begin{equation}
f_{a}\to f_{a}-T_{c}\frac{t^{2}}{\zeta^{2}}\ln{\cal \mathcal{Z}_{FSS}}(\zeta),\label{eq:mortoj}\end{equation}
 and, since $f_{a}$ is also proportional to $t^{2}$, the scaling
form for the specific heat becomes:\begin{equation}
\frac{C_{N}(T)}{C_{\infty}(T)}=c(tN^{1/2}).\label{eq:mortol}\end{equation}
Thus the specific heat curves for finite size systems, $C_{N}(T)$,
when scaled by $C_{\infty}(T)$and plotted as a function of $tN^{1/2}$,
should collapse to a single curve. The results of the Monte Carlo
simulations are consistent with this prediction (see the inset of
Fig.~(\ref{cap:Cv})). An excellent accordance between the analytic
behavior and the results of direct summation in Eq. (\ref{eq:4})
and of MC simulations is shown in Fig. \ref{cap:Sum-Sim}.

A similar scaling is observed in the susceptibility (not shown here)
and has been observed by other authors in 1D \cite{hong02} and also
in 2D and 3D \cite{herrero02}, (where $N=L^{d}$, is the number of
spins, not the linear dimension $L$ of the lattice). 

This is the expected form of scaling for a situation in which the
dimensionality is \emph{greater} than the upper critical dimension
and hyperscaling is violated \cite{binder85}. It is observed in all
these small world models for any dimension of the underlying regular
lattice \cite{hong02,herrero02}. A similar steepest descent solution
probably applies also in all these cases.

\section{Relation to Bethe lattice approach}

\label{sec:Relation-to-Bethe}
\begin{figure}
\begin{center}\includegraphics[%
  width=0.50\linewidth]{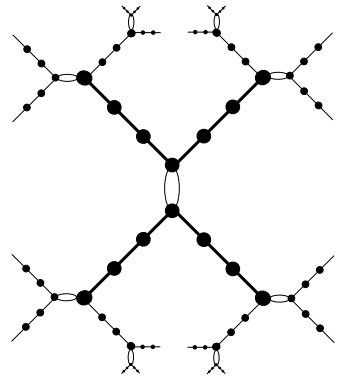}\end{center}

\caption{The Bethe lattice that describes the local environment of any spin
in the model: shortcuts (double lines) between the sites divide the
Ising chain into equal segments with $d$ (here $d=3$) chain bonds
(single lines).\label{cap:bethe-lattice}}
\end{figure}

The local environment of a spin in our
model looks like the Cayley tree in Fig.~(\ref{cap:bethe-lattice}).
The vertical links are shortcuts (of strength $I)$ and the longer
ones segments of the 1D chain, containing $d=1/2p$ links of strength
$J$. Since a shortcut from a given spin has an equal chance of linking
it to anywhere in the lattice we do not expect to find closed loops
until we go $\mathcal{O}(\ln N)$ links away. Based on this insight,
Dorogovtsev \emph{et. al} \cite{dorogovtsev02} \emph{}developed a
description of the Ising model on such lattices based on the Bethe
lattice solution. Their detailed results (namely for $T_{c}$) are
not directly applicable to our lattices (which are not maximally random
because of the strong correlation between shortcut sites). But their
insight certainly is, and, since our results are not based on the
Bethe lattice solution, and our lattice has a well defined thermodynamic
limit, it is interesting to consider the relation between the two
approaches. %

In a Bethe lattice there is only one path to link two spins, as a
result the correlation function is of 1D character. Therefore, the
correlation between any two spins $\langle\sigma_{0}\sigma_{r}\rangle$
decays exponentially with the distance at any finite temperature (even
at $T_{c}$). But the number of spins $\sigma_{r}$ at a given distance,
$r$, from a given one, $N_{s}(r)$, grows exponentially with distance,
not as a power law, as in a regular lattice. The function $\langle\sigma_{0}\sigma_{r}\rangle N_{s}(r)$
has a decay length that diverges at the ordering temperature of the
Bethe lattice. It is straightforward to derive the exact transition
temperature of a Bethe lattice of coordination $q$, $\beta_{c}J=\ln\left(q/(q-2)\right)/2$,
from this condition. 

In our model the chemical distance between two spins can be taken
as $L'+M$ where $M$ is the number of shortcuts and $L'\geq M$ the
number of 1D chain segments with $d$ bonds each, which connect these
spins. Then the 1D correlator is\begin{equation}
\langle\sigma_{0}\sigma_{(L',M)}\rangle=e^{L'd\ln t_{J}+M'\ln t_{I}}\label{eq:mortom}\end{equation}
The total number of spins with this separation from a given one, is
\begin{equation}
N_{s}(L',M)=2^{M}C_{M}^{L'},\label{eq:18}\end{equation}
therefore\begin{equation}
\langle\sigma_{0}\sigma_{(L',M)}\rangle N_{s}(L',M)=e^{-\kappa(x,T)(L'+M)},\label{eq:19}\end{equation}
where the decay constant $\kappa(x,T)$, with $x\equiv L'/(L'+M)$,
is \begin{eqnarray}
\kappa(x,T) & = & (1-x)\ln(1-x)+(2x-1)\ln(2x-1)-\nonumber \\
 &  & (1-x)\ln2-x\ln x\nonumber \\
 &  & -x\ln t_{J}^{d}-(1-x)\ln t_{I}\label{eq:morton}\end{eqnarray}
Minimizing $\kappa(x,T)$ with respect to $x$, we find that this
minimum value $\kappa(T)=\min_{x}\kappa(x,T)$ decreases with temperature
and turns zero just at $T=T_{c}$ given by Eq.(\ref{eq:13}). At any
temperature above $T_{c}$ the function in Eq.~\ref{eq:19} decays
exponentially and there is no possibility of long range order. In
the language of the Bethe lattice, the occurrence of an extra term
in the free energy below $T_{c}$, expresses the effect of boundaries,
which is never negligible, no matter what the lattice size is, when
the function in Eq.~\ref{eq:19} does not decay with distance. 

The finite size scaling for this Bethe lattice again corresponds to
the above referred situation when the hyperscaling relations are violated
and the finite size corrections are not determined by the length that
characterizes the decay of correlation functions. In fact, we have
argued above that the spin-spin correlation function decays exponentially
with the distance measured on the equivalent Bethe lattice\emph{,
i.e.} with the chemical distance on the original lattice. Thus, for
the instance of $I=J$, we have $\xi=-1/\ln t_{J}$ and\begin{equation}
\langle\sigma_{0}\sigma_{(L',M)}\rangle=e^{-\frac{L'd+M}{\xi}}.\label{eq:mortoq}\end{equation}
 It is well known that the chemical distance between any two randomly
chosen spins is of order $\mathcal{O}(\ln N)$ so the correlation
between two spins at a distance of order $N$ along the 1D chain is
at least of order $N^{-1/\xi}$; it does not decay exponentially with
$N$ as the lattice and the distance between spins grow. 

One interesting question that remains unanswered is whether one can
modify the model in order to effectively be at or below the upper
critical dimension, and therefore observe a non mean-field behavior.

\section{Conclusions}

In summary, we have been able to derive an exact solution of an Ising
model on a lattice with long range disordered interactions. This solution
expresses the free energy in terms of the density of states as a function
of two macroscopic variables of order $O(N)$, which therefore is
self-averaging. Hence no disorder averaging is required in this approach.
We obtained the thermodynamics in the $N\to\infty$ limit and also
the finite size scaling behaviour.

\section{Appendix}

The model in which the positions of the shortcut sites are randomly
chosen can be solved along the same lines of the model considered
in the main text. We denote the coordinates of the shortcut sites,
in increasing order, by $\{ q_{1},q_{2},\dots,q_{2pN}\}$ with $q_{1}>0$
(allowing for $q_{2pN}=0$) . The distances between consecutive sites
are\begin{eqnarray}
d_{i} & = & q_{i}-q_{i-1},\qquad i\neq1,\nonumber \\
d_{1} & = & N-q_{2pN}+q_{1}\label{eq:mortor}\end{eqnarray}
so that $\sum_{i=1}^{2pN}d_{i}=N$. If we choose $M$ bonds (those
for which $b=1$) from the total of $pN$, the corresponding shortcut
sites $\{ r_{1},r_{2},\dots,r_{2M}\}\subseteq\{ q_{1},q_{2},\dots,q_{2pN}\}$
will divide the lattice into $2M$ segments of lengths $l_{1},\dots,l_{2M}$
where \begin{eqnarray}
l_{i} & = & r_{i}-r_{i-1},\qquad i\neq1,\nonumber \\
l_{1} & = & N-r_{2M}+r_{1},\label{eq:mortos}\end{eqnarray}
and $\sum_{i=1}^{2M}l_{i}=N$. Then $\Omega(L,M)$ is the number of
possible choices of the $M$ bonds such that \begin{equation}
l_{2}+l_{4}+l_{6}+\dots+l_{2M}=L.\label{eq:20}\end{equation}
We define\begin{equation}
\Omega(L,M)=\Omega(M)P(L|M),\label{eq:mortoz}\end{equation}
where $\Omega(M)$ is the number of choices of $M$ shortcuts with
$b=1$ from a total of $pN$ ($\Omega(M)=C_{M}^{pN}$), and $P(L|M)$
is the probability that any such choice of $M$ bonds will select
$L$ chain bonds with $b=1.$ 

It should be stressed again that, for a given realization of disorder,
this probability must be calculated in the event space consisting
of the choices of $M$ shortcuts from the specific set of $pN$ random
shortcuts. However, it follows from the self-averaging property in
the thermodynamic limit that any statistically significant configuration
of shortcuts leads to the same probability $P(L|M)$. In that case
we can calculate it, enlarging the space of events to include \emph{all}
the configurations of shortcuts.  

We are therefore led to ask in how many ways one can choose $2M$
sites, $\{ q_{1},q_{2},\dots,q_{2M}\}\subseteq\{1,\dots,N-1\}$, such
that the sum of even lengths in this series is $l_{2}+l_{4}+l_{6}+\dots+l_{2M}=l$.
We have seen above that it is given by\begin{equation}
C_{M}^{N-L}C_{M-1}^{L-1}\label{eq:mo1}\end{equation}
Since the total number of such choices is $C_{2M}^{N}$, we have \begin{equation}
P(l|s)=C_{M}^{N-L}C_{M-1}^{L-1}/C_{2M}^{N},\label{eq:mo2}\end{equation}
so that 

\begin{equation}
\Omega(L,M)=C_{M}^{pN}C_{M}^{N-L}C_{M-1}^{L-1}/C_{2M}^{N}.\label{eq:mo3}\end{equation}
With the same definitions as above we get\begin{eqnarray}
s(l,n) & = & -pn(\ln n/4)-p(1-n)\ln(1-n)\nonumber \\
 & + & (1-2pn)\ln(1-2pn)\nonumber \\
 & - & (1-l-pn)\ln(1-l-pn)\nonumber \\
 & - & (l-pn)\ln(l-pn)\nonumber \\
 & + & (1-l)\ln(1-l)+l\ln l,\label{eq:entropyB}\end{eqnarray}
 The equation for the transition temperature is \begin{equation}
t_{J}(1+4pt_{I})=1\label{eq:ssss}\end{equation}
with the characteristic limits \begin{eqnarray}
T_{c} & = & \frac{2J}{\ln(1/2p)},\qquad p\ll1,\quad I>\frac{J}{\ln(1/2p)},\nonumber \\
T_{c} & = & \frac{2J}{\ln\left(T_{c}/2pI\right)}\qquad pI\ll J.\label{eq:dddd}\end{eqnarray}
The resulting thermodynamic behavior is essentially the same as in
the model considered in the main text. 

\begin{acknowledgments}
We thank Sergei Dorogovtsev for pointing out the relation of our solution
to the Bethe lattice. J. Viana Lopes was supported by FCT grant (Portugal)
No. SFRH/BD/1261/2000 and Raul Toral by MCYT (Spain) and FEDER through
projects BFM2000-1108 and BFM2001-0341-C02-01.
\end{acknowledgments}
\bibliographystyle{apsrev}
\bibliography{sw}

\end{document}